\documentclass[conference]{IEEEtran}

\usepackage{epsfig,endnotes}
\usepackage[cmex10]{amsmath}
\usepackage{algorithmic}

\usepackage{array}
\usepackage{fixltx2e}

\usepackage{mdwmath}
\usepackage{amsmath}
\usepackage{multirow}
\usepackage{mathtools}

\usepackage{mdwtab}
\usepackage[flushleft]{threeparttable}
\usepackage{stfloats}
\usepackage[ruled,vlined]{algorithm2e}

\usepackage{slashbox}
\usepackage{caption}
\usepackage{subcaption}

\usepackage{threeparttable}

\usepackage{cite}

\makeatletter
 \let\@copyrightspace\relax
\makeatother

\usepackage{epstopdf}

\usepackage[utf8x]{inputenc}

\begin{document}
%
\date{}
\title{EchoIA: A Human-Centered Implicit Authentication Leveraging User Feedback}



\author{
\IEEEauthorblockN{Yingyuan Yang\IEEEauthorrefmark{1}, Jiangnan Li\IEEEauthorrefmark{2}, Sunshin Lee\IEEEauthorrefmark{1}, Dawei Li\IEEEauthorrefmark{3},  and Jinyuan Sun\IEEEauthorrefmark{2}\\}
\IEEEauthorblockA{\IEEEauthorrefmark{1}University of Illinois, Springfield, IL, 62629 USA Email: \{yyang260,slee675\}@uis.edu\\}
\IEEEauthorblockA{\IEEEauthorrefmark{2}University of Tennessee, Knoxville, TN, 37996 USA Email: \{jli103,jysun\}@utk.edu\\}
\IEEEauthorblockA{\IEEEauthorrefmark{3}Department of Computer Science, Montclair State University, Montclair, NJ, 07043, USA. Email: dawei.li@montclair.edu}
\\\vspace*{-1.2cm}
}


%


\maketitle
\pagestyle{plain}

\begin{abstract}
Implicit authentication (IA) transparently authenticates users by utilizing their behavioral data sampled from various sensors. Identifying the illegitimate user through constantly analyzing current users' behavior, IA adds another layer of protection to the smart device. Due to the diversity of human behavior, existing research tends to utilize multiple features to identify users, which is less efficient. Irrelevant features may increase the system delay and reduce the authentication accuracy. However, dynamically choosing the best suitable features for each user (personal features) requires a massive calculation, making it infeasible in the real environment. In this paper, we propose EchoIA to find personal features with a small amount of calculation by leveraging user feedback derived from the correct rate of inputted passwords. By analyzing the feedback, EchoIA can deduce the true identities of current users and achieve a human-centered implicit authentication. In the authentication phase, our approach maintains transparency, which is the major advantage of IA. In the past two years, we conducted a comprehensive experiment to evaluate EchoIA. We compared it with four state-of-the-art IA schemes in the aspect of authentication accuracy and efficiency. The experiment results show that EchoIA has better authentication accuracy (93\%) and less energy consumption (23-hour battery lifetimes) than other IA schemes.\\

\end{abstract}


%

\section{Introduction}
Recent years have witnessed the rapid growth of smart technologies such as the smartphone, smartglasses, and smartwatch. On one hand, people rely heavily on smart devices to share information and to gain services, which become primary elements of our daily life. On the other hand, the security problem raised by smart devices becomes more important than ever before. One of the most important issues is user authentication.

To identify users, most of the existing systems use explicit approaches (explicit authentication), such as passwords, PINs, and draw-patterns. However, explicit authentication requires user-system interaction, which could be frustrating especially when the users possess many different passwords. A recent survey \cite{WhatIS} shows 3\% percent of people forget a password at least once a week. Explicit authentication can also be circumvented and be broken \cite{p40}. Researchers begin to study new authentication methods to enhance explicit authentication.

Utilizing sensors' data sampled by the smart device, implicit authentication (IA) transparently identifies users by constantly comparing current users' behavioral data with historical legitimate users' behavioral data. The comparing process, or classification process, is usually achieved by using various machine learning models, e.g., SVM. Since users do not need to interact with the system in the authentication, IA can be seamlessly applied to various authentication systems, which adds another layer of protection to the smart device. During the usage, if suspicious behaviors were detected, IA will lock the device and ask users to perform a multi-factor authentication, e.g., inputting passwords. From the users' aspect, due to IA's transparency, they will not be able to notice IA until the device has been locked. In addition, implicit authentication does not require user-system interaction, which releases users from tedious passwords inputting and the burden of memorizing the passwords.

In implicit authentication (IA), the features used for the authentication are predefined by the system, which will not be able to change during the usage \cite{shen2018performance,castelluccia2017towards}. To achieve better coverage in user authentication, the existing approaches tend to use multiple features \cite{yang2020bubblemap,bello2020machine,xu2020touchpass}, such as location, touch, and acceleration for user identification purposes. However, for a specific user, only a small number of the features (personal features) in the total feature set is needed. Irrelevant features not only reduce the system's efficiency, but decrease the authentication accuracy as well. Nevertheless, due to the high complexity of human behavior, it requires a massive calculation to derive personal features, which is infeasible in practice \cite{yang2016personaia}. Hence, finding a suitable scheme that dynamically derives personal features is a critical step for IA implementation.


\begin{figure*}[htb]
\vspace*{-0.3cm}
\centering
  \begin{subfigure}[b]{3.33in}
    \centering
    \includegraphics[width=3.33in, height=1.3in]{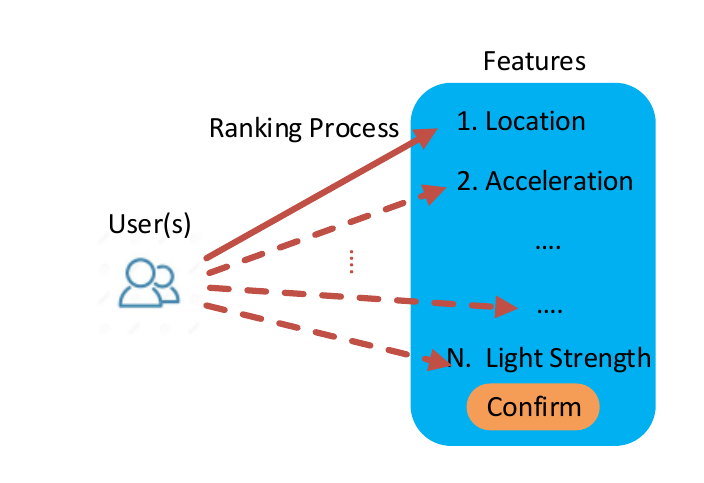}
    \caption{}\label{fig:Sysa}
  \end{subfigure}%
  \begin{subfigure}[b]{3.67in}
    \centering
    \includegraphics[width=3.67in, height=1.3in]{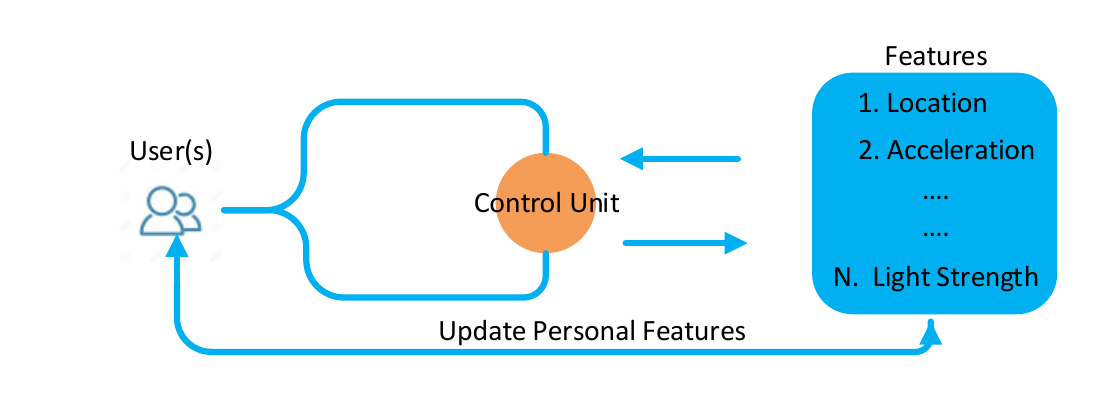}
    \caption{}\label{fig:Sysb}
  \end{subfigure}%
  \caption{EchoIA overview. (a) The \emph{Initialization} phase. (b) The \emph{Authentication} phase.}
  \label{fig:system}
  \vspace*{-0.5cm}
\end{figure*}

In this work, we introduce a human-centered implicit authentication (EchoIA), which utilizes user feedback to pinpoint personal features during the usage with a small amount of calculation. By comparing with current IA schemes, the experiment shows that our method can significantly improve the authentication accuracy of traditional IA. In addition, the proposed method is lightweight, which can be easily embedded into existing IA systems as an add-on to achieve efficient authentication. As far as we know, we are the first group, which utilizes user feedback to improve the authentication accuracy and energy efficiency in IA.

The major advantage of implicit authentication (IA) is its transparency, which releases users from the tedious authentication process. However, it is difficult to gather user feedback in a transparent environment, since directly asking users' input will break the transparency. Even though we could have various user feedback, pinpointing the best suitable features is also challenging. To this end, we propose a method that utilizes the correct rate of inputted passwords to implicitly collect user feedback and find personal features. In implicit authentication, the system will lock the device and deem current users illegitimate when their behavior mismatch legitimate users' historical behavior. For legitimate users, they may be locked out due to the misidentification caused by using unsuitable features, but they can input a correct password to unlock the device. For illegitimate users, they may also input a correct password to unlock the device after several attempts, but their correct rate of inputted passwords will be lower than legitimate users'. EchoIA can utilize the correct rate of inputted passwords to deduce current users' true identities and further adjust feature sets to better match users' behavior. The detailed procedure is discussed in Section II.

This paper makes the following contributions:

$\bullet$ We proposed EchoIA to find the best suitable features (personal features) for each user by utilizing user feedback. EchoIA maintains the transparency of implicit authentication, while can choose personal features for legitimate users based on their recent behavior.

$\bullet$ We implemented EchoIA in a real environment by using the Android system and multiple servers. To evaluate the proposed method, we also implemented four state-of-the-art implicit authentication schemes.

$\bullet$ We collected users' behavioral data in the past two years. In addition, we evaluated the proposed method in the aspect of authentication accuracy, computational efficiency, and energy efficiency. In the experiment, EchoIA has a better authentication accuracy and a lower energy cost.

\section{The System Overview \label{sysOverview}}

Implicit authentication (IA) identifies users by constantly comparing current users' behavioral data with legitimate users' historical behavioral data. If current users' behavioral data is different, IA will lock the device and ask the users to input passwords. Due to the noise and behavioral change, it is common that legitimate users are falsely blocked by the device \cite{yang2016personaia}. In implicit authentication, if users input correct passwords, they can continue to use the device, where the setting of the original IA system will not change. In EchoIA, however, if users input correct passwords and prove their identities, the system will not only unlock the device, but may also adjust the features to align with legitimate users' behavior.

As shown in Fig. \ref{fig:system}, EchoIA contains two phases, \emph{Initialization} and \emph{Authentication}. The \emph{Initialization} phase only takes place for the first time of the usage. Meanwhile, we assume the users at the \emph{Initialization} phase are legitimate. In the \emph{Initialization} phase, candidate features are sent to legitimate users, who may rank the features based on their behavior. The combination of top-ranking features and system-default features is used as personal features to identify users. Note that the personal features are dynamically adjusted according to different users and devices. In the \emph{Authentication} phase, EchoIA utilizes the correct rate of inputted passwords to adjust personal features. Specifically, user feedback is implicitly obtained through the process of inputting passwords (Section \ref{initialization}), which keeps IA's transparency. However, to prevent illegitimate users take advantage of the system, it only updates personal features after users entered a correct PIN number, which must be different from passwords used for unlocking the device. A secured channel is established to transmit data between client and server. We adopted Wind Vane module \cite{yang2016personaia} to optimize the data transmission efficiency.

From the system's point of view, inputting incorrect passwords will enhance its confidence in using existing personal features; inputting correct passwords will reduce its confidence in using existing personal features and will encourage it to choose different features. Since most of the time the system is running at the \emph{Authentication} phase, the users will not be able to notice the existence of EchoIA during the usage. The following sections will discuss the detail of the \emph{Initialization} phase and the \emph{Authentication} phase.

\subsection{The Initialization Phase \label{initialization}}
As its name suggests, the \emph{Initialization} phase mainly focuses on initializing personal features and associated system settings. The users will spend a short time in this phase in order to help the system to prepare the authentication.

As shown in Fig. \ref{fig:system} (a), at the \emph{Initialization} phase, EchoIA will send a message contains all candidate features to the users. Based on their own behavior, the users will rank candidate features, where the result will be sent back to the remote servers for further processing. For each feature, there is an associated weight parameter, which will be initialized at this phase. The total available features in the smart device are $F$. Note that the elements in $F$ are various for different devices, and can be updated during the usage once new features are introduced.

\begin{equation}
\label{fun1}
\begin{split}
F=\{f_1,f_2,f_3,...,f_n\}
\end{split}
\end{equation}

To ensure reliability, some of the features are system preserved, which is not shown in $F$. For example, a touch trajectory feature is preserved since it has high accuracy when identifying most of the users. For each feature in $F$, the corresponding weight is predefined in $W$.

\begin{equation}
\label{fun2}
\begin{split}
W=\{w_1,w_2,w_3,...,w_n\}
\end{split}
\end{equation}

The users may rank the features based on their routine. To this end, EchoIA will renew the weight for each feature based on the users' ranking.

\begin{equation}
\label{fun3}
\begin{split}
w_n=\frac{1}{r_n},
\end{split}
\end{equation}
where $w_n$ is the weight of the $n$th feature; and $r_n$ is the associated ranking of the feature. The top-ranking features only contain a part of elements in $F$ and is dynamically changed during the usage. For example, at some moments, the top-ranking features, $F_{top}$, may only contain 5 different features $F_{top}=\{f_2,f_4,f_5,f_7,f_8\} \subset F$. Personal features in this example will have both $F_{top}$ and system reserved features.

In real usage, users may change their behavior, which is common in practice. To better identify the users, the system also needs to adjust personal features according to the behavioral change. The details of adjusting personal features will be discussed in the next section.

\subsection{The Authentication Phase}
To dynamically adjust personal features, the system can directly send the message to users to request new features, but this approach will break IA's transparency. In addition, it is difficult for the system to decide the ``right time" to send the request, since the system will not know the behavioral change unless it analyzed the data. In EchoIA, instead of analyzing users' behavioral data, the system leverages the correct rate of inputted passwords to dynamically adjust personal features.

At the \emph{Authentication} phase, as shown in Fig. \ref{fig:system} (b), the system will reduce the weights of each feature in $F_{top}$ if users input correct passwords. Since users only need to input passwords when IA locks the device, correct passwords indicate current users have a large chance of being legitimate. Similarly, the system will increase the weight of each feature in $F_{top}$ if users input incorrect passwords. The new weight is updated by $\delta_n$.

\begin{equation}
\label{fun3}
\begin{split}
\delta_n=\delta_n^{(I)}-\delta_n^{(C)},
\end{split}
\end{equation}
where $\delta_n^{(I)}$ is the amount of weight increased for the feature $n$ in $F_{top}$; $\delta_n^{(C)}$ is the amount of weight decreased for the feature $n$ in $F_{top}$. The new weight of the feature $n$ is calculated by $w_n+\delta_n$.

In EchoIA, a predefined threshold $\Delta$ is used to measure the significance of the weight change. In practice, we choose $\Delta$ by using k-fold cross-validation. If the change is significant, $\delta_n > \Delta$, the system will challenge the users to input a PIN number, which is the number different from passwords used to unlock the device. The users can choose the PIN number at the \emph{Initialization} phase. If the users type a correct PIN and agree with personal features' change, EchoIA will update $F_{top}$ and personal features according to the new weights in $F$. In this process, decayed features will be replaced by new features. The system will use the updated personal features to identify the users until $\delta_n > \Delta$ again. We adopted the support vector machine (SVM) to achieve the user classification and to identify legitimate users. The parameters in the model are optimized by using k-fold cross-validation.

\section{Implementation}
\begin{figure}[htb]
\centering
\vspace{-0.4cm}
\hspace*{-0.1cm}
\includegraphics[width=3.3in,height=1.5in]{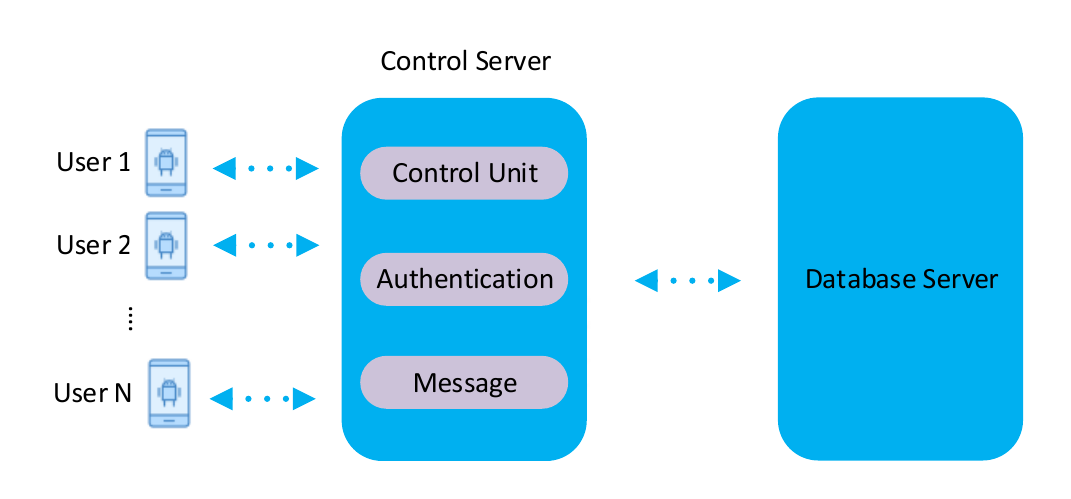}
\caption{The system architecture of EchoIA.}
\vspace{-0.3cm}
\label{fig:systemChart}
\end{figure}

We implemented EchoIA by using the Android system and multiple servers. The system architecture is shown in Fig. \ref{fig:systemChart}. The data collection is achieved at the user-end, in which an application is created to collect users' behavioral data. The user-end application constantly samples users' behavioral data from various sensors and sends it to the Control Server by using a secured channel. As shown in Fig. \ref{fig:systemChart}, multiple users can connect with the Control Server at the same time. The Control Server contains three main components, Control Unit, Authentication Unit, and Message Unit. As mentioned in section \ref{sysOverview}, the Control Unit is responsible for updating the weight parameter associated with each feature. The Authentication Unit leverages implicit authentication to constantly monitor users' behavior and compare it with legitimate users' historical behavior. In order to compare EchoIA with other IA schemes, we also implemented four state-of-the-art IA schemes in the Authentication Unit, called Shi-IA \cite{p9}, Multi-Sensor-IA \cite{yang2020bubblemap}, Gait-IA \cite{frank2010activity}, and SilentSense-IA \cite{bo2013silentsense}. Finally, the Message Unit is used to communicate with users during the \emph{Initialization} phase and the \emph{Authentication} phase. All users' data is formatted and stored in the Database Server.

Written by java, the user-end application is installed in two types of smart devices, Motorola G2 with the Android system version of 6.0 and Samsung Galaxy S10 with the Android version of 9.0. The Motorola G2 has 1GB memory, 8GB local storage, a Quad-core 1.2 GHz CPU, and Adreno 305 GPU. The Samsung Galaxy S10 has 8GB memory, 128GB local storage, an Octa-core 1.95 GHz CPU, and Adreno 640 GPU. The Control Server has a Qual-core 2.4GHz CPU, 16GB memory, 500GB local storage, and an NVIDIA GeForce GTX 660M GPU. We adopted Google Firebase as the remote database server, which is a NoSQL database and can easily store more than thousands of users' encrypted data. A program written by JavaScript is used to achieve data transmission.

\begin{figure}[htb]
\centering
\vspace{-0.4cm}
\hspace{-0.72in}
  \begin{subfigure}[b]{1.7in}
    \centering
    \includegraphics[width=2.3in, height=2.2in]{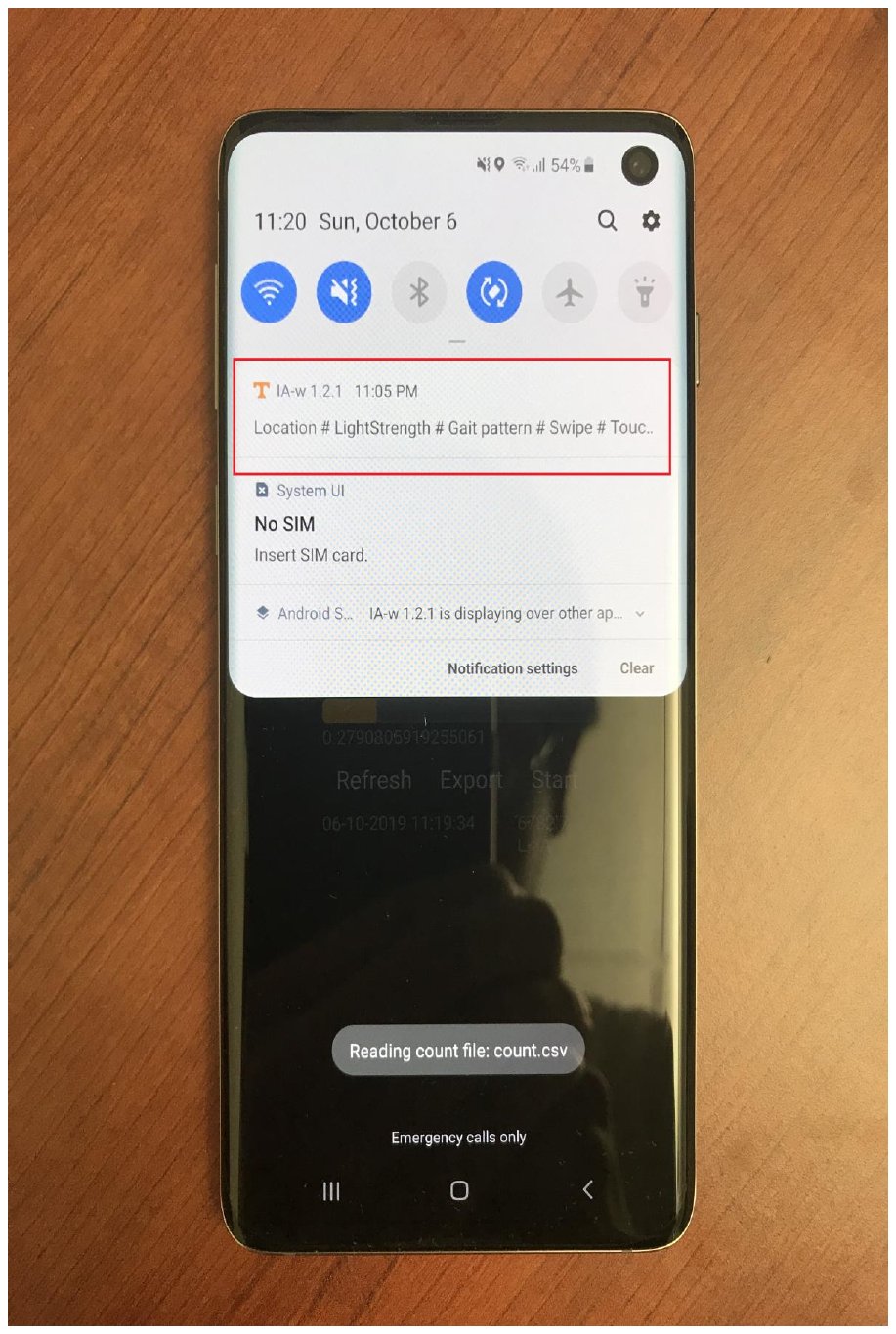}
    \caption{}\label{fig:realA}
  \end{subfigure}%
  \begin{subfigure}[b]{1.7in}
    \centering
    \includegraphics[width=2.3in, height=2.2in]{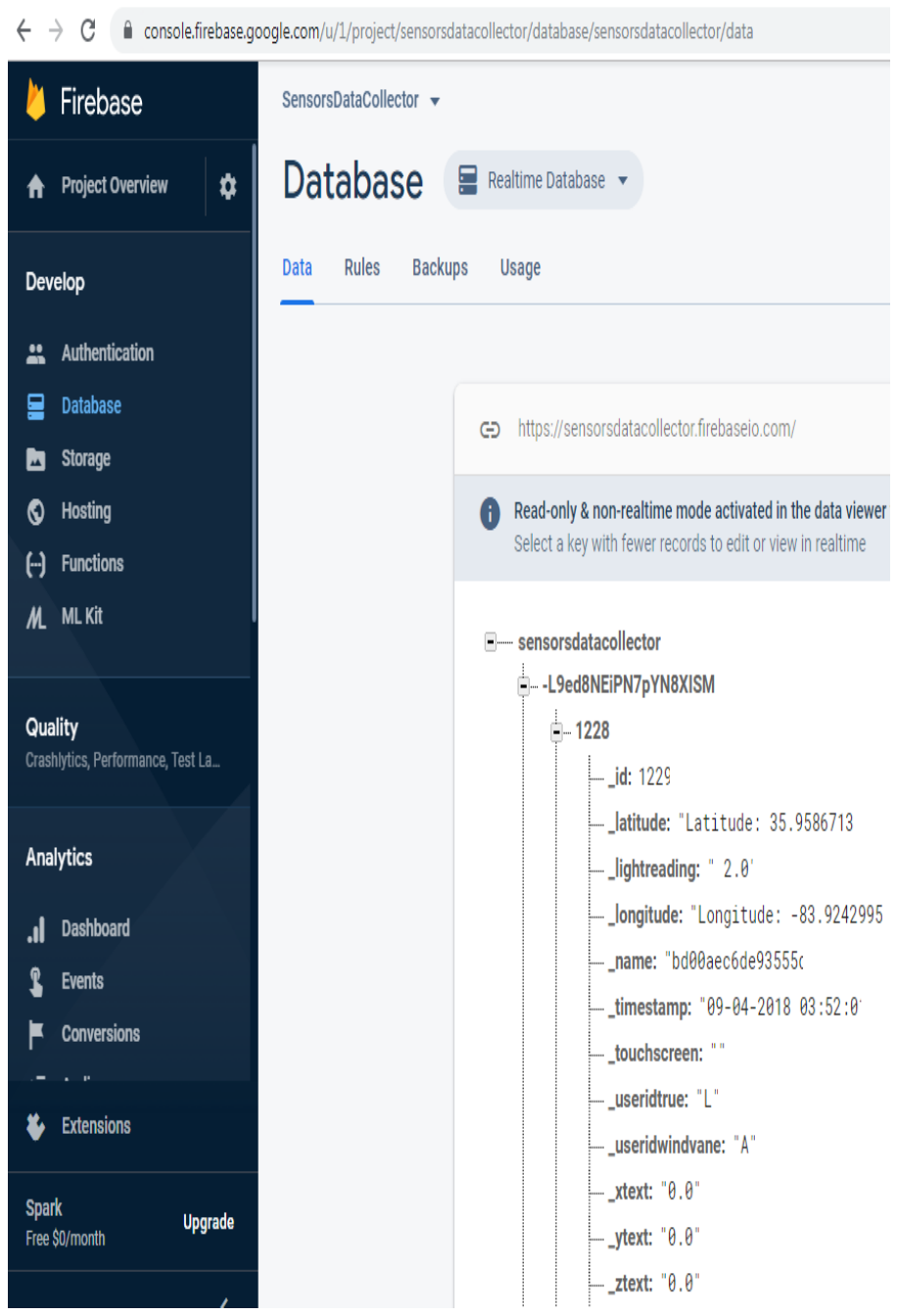}
    \caption{}\label{fig:realB}
  \end{subfigure}%
  \caption{EchoIA implementation. (a) The user-end application. (b) The server-end database.}
  \label{fig:implementation}
\end{figure}

The user-end application and the server-end database are shown in Fig. \ref{fig:implementation} (a) and (b) respectively. The red rectangle in Fig. \ref{fig:implementation} (a) indicates candidate features sent from the Control Server. The user interface is shown under the message pop-up, in which the legitimate user can rank candidate features by using the GUI provided by the application. The user-end application also shows the behavioral data associated with each feature, e.g., light strength 2.0, longitude -83.92, and latitude 35.95. The sampling frequency is dynamically controlled for energy-saving purposes \cite{yang2016personaia}. As shown in Fig. \ref{fig:implementation} (b), users' data is stored in the database server as JSON files, in each of which contains the data sampled from a smart device. For example, Fig. \ref{fig:implementation} (b) shows the data sampled from a smart device with an ID of L9ed8NEiPN7pYN8XISM. The data is stored based on the time stamp. Note that the sensitive data is hashed before send to the Control Server.

\vspace{-0.3cm}

\section{Evaluation}
In the real experiment, we tracked the usage of 17 participants during the past two years. Each user was in turn selected as the legitimate user, while another user was deemed as illegitimate users. Illegitimate users were required to use the device more than 10\% of the total usage time. Utilizing our system, we gathered rich information from all users under different environments. In the experiment, we mainly use the following 12 features to achieve user authentication: accelerometer, orientation, magnetometer, gyroscope, touch, light, pressure, temperature, GPS, microphone, battery usage, and wifi status.

To compare EchoIA with other state-of-the-art IA schemes \cite{khan2014comparative}, we implemented Shi-IA \cite{p9}, Multi-Sensor-IA \cite{yang2020bubblemap}, Gait-IA \cite{frank2010activity}, and SilentSense-IA \cite{bo2013silentsense}. We used the recommended settings of the original papers \cite{p9,yang2020bubblemap,frank2010activity,bo2013silentsense} in the experiment. In addition, the feature selection strictly follows the descriptions of the original works, while the parameters were optimized by using k-fold cross-validation. Finally, we tested the performance of different schemes by using the same data.

In the experiment, we evaluated the authentication accuracy, CPU utilization, memory utilization, and energy consumption for each IA schemes, including EchoIA. The experiment details are described in the following sections.

\subsection{Authentication Accuracy}
\begin{figure}[htb]
\centering
\vspace{-0.4cm}
\includegraphics[width=3.8in,height=1.7in]{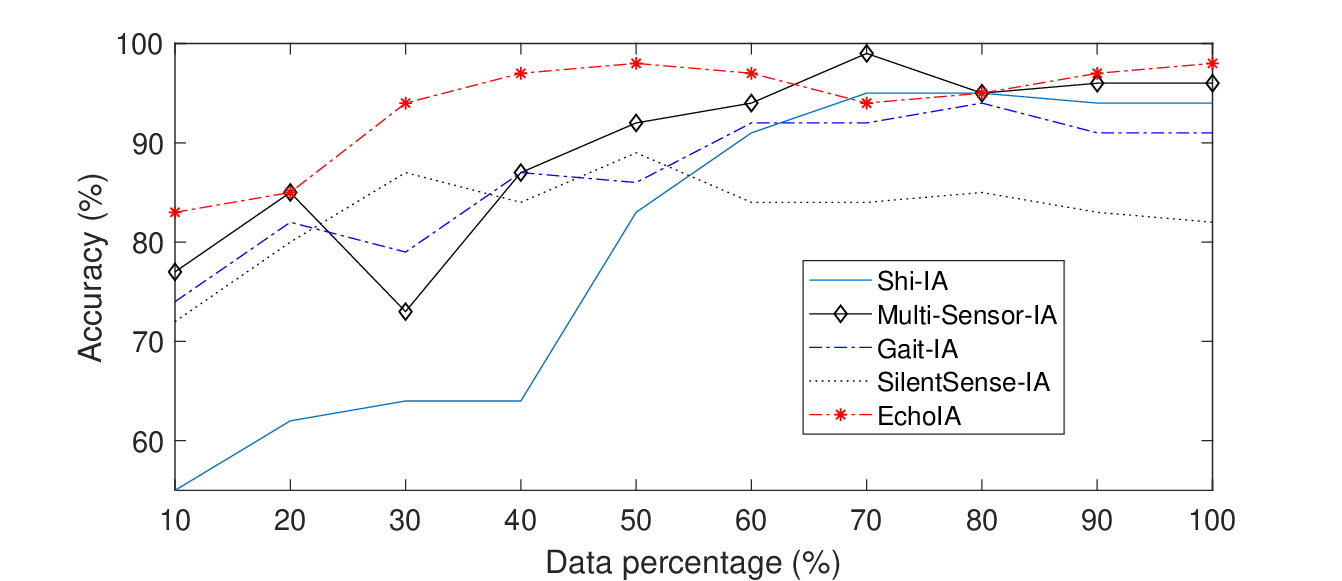}
\caption{Accuracies for different IA schemes.}
\vspace{-0.3cm}
\label{fig:AccScheme}
\end{figure}

We first compared the authentication accuracy of different IA schemes, where results are shown in Fig. \ref{fig:AccScheme}. In the figure, users' data was divided into ten parts based on the timeline. The first part of the data was used to train the models \footnotemark. \footnotetext{The first parts contain 15\% of the total data since the training data of this size optimizes the authentication accuracy across all five IA schemes.} We then calculated the authentication accuracies for different schemes by using the rest parts of the data. The authentication accuracy was calculated by $ACC=\frac{TR+TA}{TR+TA+FR+FA}$, where true accept (TA) indicates the legitimate user has been correctly identified; the false reject (FR) indicates the legitimate user has been incorrectly identified to be the illegitimate user; the true reject (TR) indicates the illegitimate user has been correctly identified; the false accept (FA) indicates the illegitimate user has been incorrectly identified to be the legitimate user. We also adopted the retraining techniques discussed in our previous work \cite{p67} to improve the authentication accuracy for all five schemes.

EchoIA has the highest authentication accuracy in most of the tests, as shown in Fig. \ref{fig:AccScheme}. Multi-Sensor-IA also has a high authentication accuracy compared to other IA schemes. In the experiment, most of the schemes reach to more than 90\% accuracy after using 80\% of the data except SilentSense-IA.

\begin{figure}[htb]
\centering
\vspace*{-0.4cm}
\includegraphics[width=3.8in,height=1.3in]{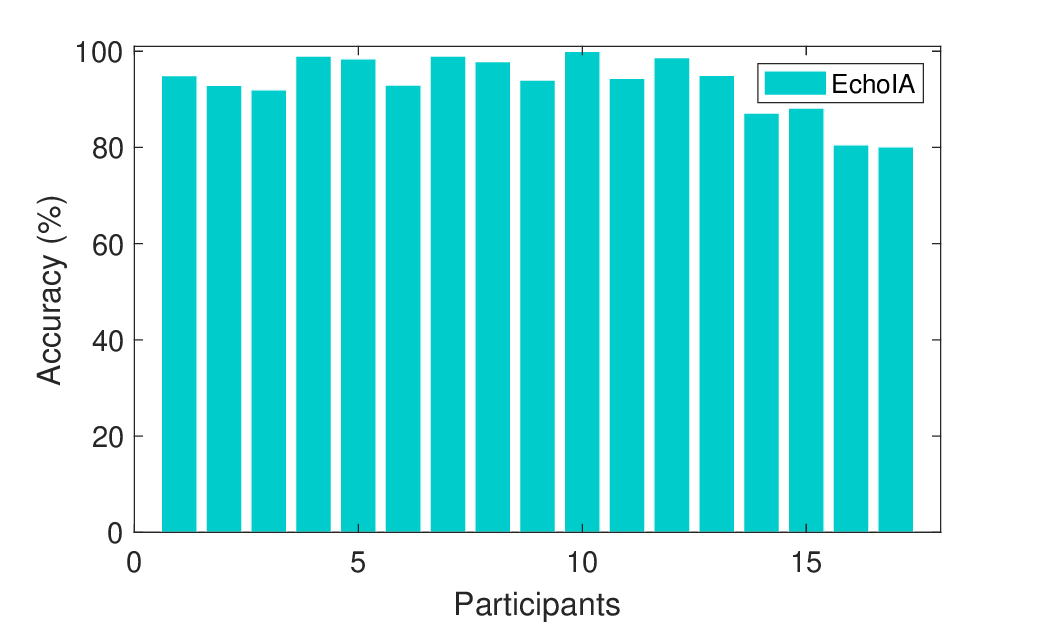}
\caption{The accuracy for each user.}
\vspace{-0.4cm}
\label{fig:AccIndivisual}
\end{figure}

In EchoIA, we calculated an average authentication accuracy for 17 users by utilizing all the data spanned two years. The result is shown in Fig. \ref{fig:AccIndivisual}, where the average accuracy across all users is 93.23\%.

\subsection{CPU Utilization and Memory Utilization}

\begin{figure}[htb]
\centering
\hspace{-0.13in}
  \begin{subfigure}[b]{1.7in}
    \centering
    \includegraphics[width=1.89in, height=1.65in]{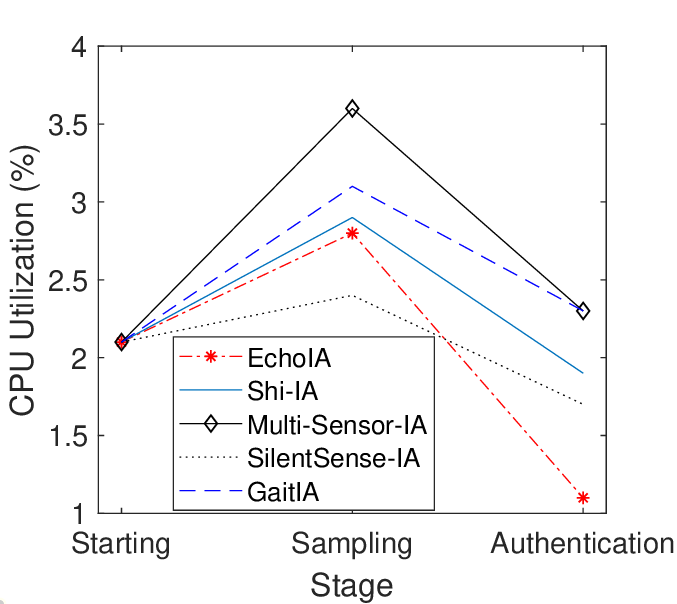}
    \caption{}\label{fig:realA}
  \end{subfigure}%
  \hspace{0.04in}
  \begin{subfigure}[b]{1.7in}
    \centering
    \includegraphics[width=1.89in, height=1.65in]{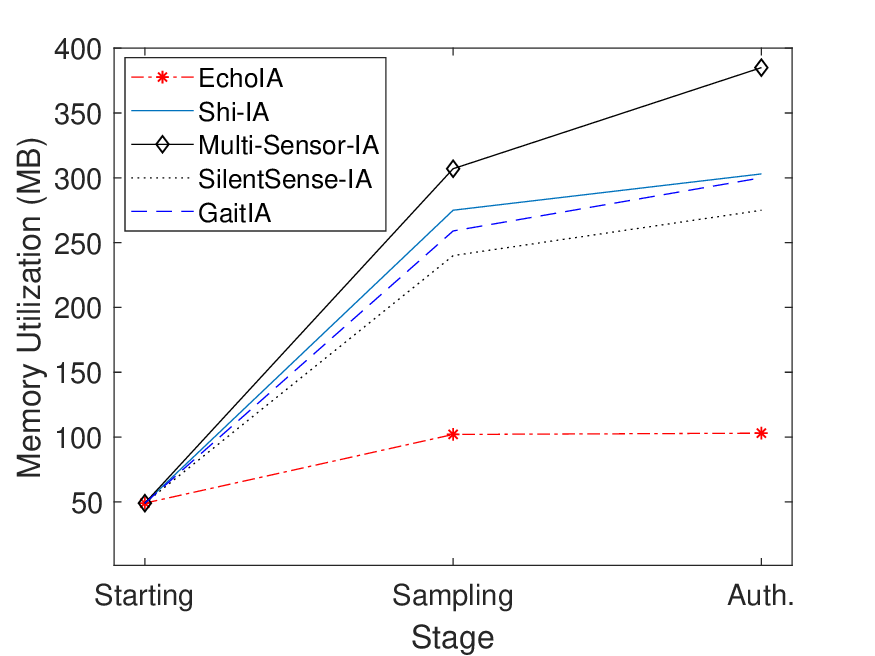}
    \caption{}\label{fig:realB}
  \end{subfigure}%
  \caption{CPU Utilization and Memory Utilization (User-End). (a) CPU utilization. (b) Memory utilization.
  \footnotesize
  *At the Starting stage, the user-end application and services begin to launch. At the Sampling stage, sensors' data is sampled and periodically uploaded to the server. The \emph{Initialization} phase is at the Sampling stage. At the authentication stage, utilizing the model returned from the server, the current user will be classified into two categories, legitimate or illegitimate. The \emph{Authentication} phase is at this stage.}
  \label{fig:CPUandMemory}
\end{figure}

In addition, we evaluated the CPU usage and memory usage for different schemes. In the experiment, we recorded the CPU utilization of different IA schemes at three stages, Start, Sampling, and Authentication. The experiment results are shown in Fig. \ref{fig:CPUandMemory} (a). At the Sampling stage, the CPU utilization of EchoIA is the second-lowest for all five schemes. At the Authentication stage, the CPU utilization of EchoIA is the lowest among all the schemes. Since most of the time the users are at the Authentication stage, the total amount of CPU utilization of EchoIA is the smallest for all the schemes. In the experiment, Multi-Sensor-IA has the highest CPU utilization, but it also has a high authentication accuracy similar to EchoIA.

We recorded the memory utilization of various schemes at different stages. The result is shown in Fig. \ref{fig:CPUandMemory} (b), in which the EchoIA has the lowest memory utilization among all the five schemes. Since EchoIA only uses a small portion of features to train the model and to authenticate users, the total amount of memory used to store the data is smaller than other schemes. As shown in Fig. \ref{fig:CPUandMemory} (b), the memory utilization of Multi-Sensor-IA is the highest since it uses all the features and associated sensors' data on the device.

\begin{figure}[htb]
\vspace*{-0.4cm}
\centering
  \begin{subfigure}[b]{1.7in}
    \centering
    \includegraphics[width=1.9in, height=1.4in]{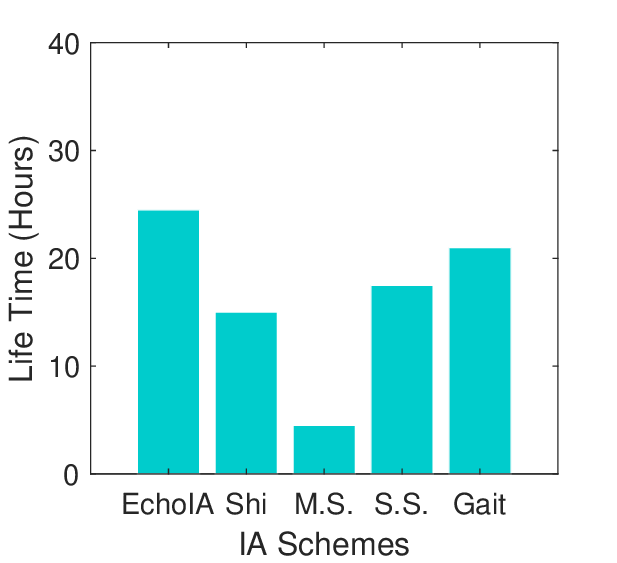}
    \caption{}\label{fig:realA}
  \end{subfigure}%
  \hspace{0.02in}
  \begin{subfigure}[b]{1.7in}
    \centering
    \includegraphics[width=1.9in, height=1.4in]{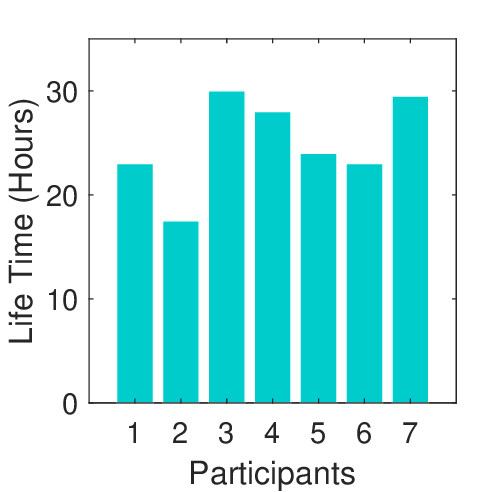}
    \caption{}\label{fig:realB}
  \end{subfigure}%
  \caption{Battery Utilization. (a) Battery utilization of different schemes. M.S. denotes Multi-Sensor-IA. S.S. denotes SilentSense-IA. (b) Battery utilization of different participants.}
  \label{fig:Battery}
  \vspace*{-0.4cm}
\end{figure}

In the experiment, we evaluated the battery consumption for different schemes. The result is shown in Fig. \ref{fig:Battery}. We measure the battery usage of different schemes by calculating the average working hours of battery after fully charged. As shown in Fig. \ref{fig:Battery} (a), EchoIA has the longest battery lifetime, 23 hours on average. The Multi-Sensor-IA has the shortest battery lifetime, four hours on average.

We also calculated the average battery lifetime of each participant by using EchoIA. Fig. \ref{fig:Battery} (b) shows the battery lifetime derived from seven different participants' data randomly selected from 17 users. There are large differences in the battery utilization between users. As shown in the figure, participant 2 has the shortest battery lifetime, which is 16 hours. Participant 7, however, has the longest battery lifetime, which is 30 hours.

\subsection{Energy Consumption of the User-End Application}

We also tracked the performance of the user-end application. In the experiment, we compared EchoIA with popular applications, such as Instagram, Facebook, Twitter, eBay, and LinkedIn. We continuously tracked CPU utilization and memory utilization for different applications during the usage. The result is shown in Table \ref{tbEnergyEfficiency}. Please note that in Fig. \ref{fig:CPUandMemory}, we calculated the CPU and battery utilization only based on the data at the sampling and authentication stages. In Table \ref{tbEnergyEfficiency}, we also gathered data from other stages, e.g., switching to a different application.

\begin{table}[!ht]
\footnotesize
\renewcommand{\arraystretch}{1.3}
\caption{CPU and Memory Consumption}
\centering
\begin{center}
\begin{tabular}{|c|c|c|c|c|c|c|}
\hline
    &Insta.*&F.B.&Twit.&L.In&eBay&EchoIA.\\
\hline
CPU Avg\%&7.1&11.6&6.0&5.9&5.3&1.3\\
\hline
CPU Max\%&10.0&16.5&10.3&9.6&8.6&3.9\\
\hline
Mem. &&&&&&\\
Max(MB)&121.2&157.0&101.4&138.2&111.6&103\\
\hline

\hline
\end{tabular}
\end{center}
\begin{tablenotes}
      \footnotesize
      \item *The abbreviation Insta. is for Instagram, F.B. for Facebook, Twit. for Twitter, L.In for LinkedIn.
\end{tablenotes}
\label{tbEnergyEfficiency}
\end{table}

Table \ref{tbEnergyEfficiency} shows the average and maximum CPU consumption for each application. EchoIA has the lowest average CPU consumption compared to other applications, which is 1.3\%. Similarly, EchoIA also has the lowest maximum CPU consumption, which is 3.9\%. EchoIA also consumes a small amount of memory in real usage, which only occupies a maximum of 103MB memory.

\section{Related Work}
To improve explicit authentication mechanisms such as PIN and passlocks, various implicit authentication schemes have been proposed as secondary authentication mechanisms \cite{ravi2005activity,el2021implicit,yang2020dynamic,cheung2020context,zhu2020espialcog,xu2020touchpass,p9,yang2020bubblemap,frank2010activity,bo2013silentsense,wei2020privacy}. Among them, leveraging different features, Shi scheme \cite{p9}, Multi-Sensor scheme \cite{yang2020bubblemap}, Gait scheme \cite{frank2010activity}, and SilentSense scheme \cite{bo2013silentsense} are four different schemes that represent four research directions of state-of-the-art implicit authentications \cite{khan2014comparative,mehrabi2020implicit}. In addition, current implicit authentication research tends to adopt all the available features to achieve better authentication accuracy \cite{yang2020bubblemap,yang2016personaia,yang2017energy}. On one hand, due to the high complexity of users' behavior, utilizing only a specific behavior metric is not sufficient to identify them in practice. On the other hand, to identify a specific user, only a small portion of the total behavior metrics is needed \cite{xu2020touchpass,khan2014comparative,yang2017energy,p9,p9,yang2020bubblemap}. Reducing the number of features can also exponentially decrease the system's energy and time consumption \cite{bishop2006pattern}. However, to find personal features requires additional calculation \cite{yang2020bubblemap,yang2016personaia}, which increases time and energy consumption. Leveraging user feedback, EchoIA can select the best suitable features for different users. During the usage, the legitimate users can also notify the system to update personal features if their behavior changed, e.g., injury.

Most of the existing research in implicit authentication utilizes the support vector machine (SVM) \cite{karanikiotis2020continuous,bo2013silentsense,yang2020bubblemap,shi2021fine,abuhamad2020sensor} to identify users. Other classifiers, e.g., Gaussian mixture model (GMM), have also been used in implicit authentication \cite{p9}. In EchoIA, we adopted SVM to achieve user authentication, but our approach is also compatible with other classifiers, such as kNN, LDA topic model, and GMM \cite{bishop2006pattern}.

\section{Conclusion}
We proposed EchoIA to find the best suitable features (personal features) for different users by utilizing user feedback. To achieve better coverage, the existing works in implicit authentication tend to use many different features to identify users, which is less efficient and may decrease the authentication accuracy. Without using additional calculations, it is difficult to dynamically choose personal features due to the transparency of IA. Leveraging the correct rate of inputted passwords, EchoIA implicitly gathers user feedback to choose personal features, while maintaining the transparency of IA. To evaluate the proposed method, we implemented EchoIA and four state-of-the-art IA schemes by using the Android system and multiple servers. The results show that EchoIA has better authentication accuracy (93\%) and less energy consumption (23-hour battery lifetimes) than other IA schemes. In the future, to benefit associated research, we will share the system's source code, parameter setting, and dataset on our lab website.

\section{Acknowledgement}
This work was partially supported by the US National Science Foundation (NSF) under grant CNS-1422665 and the Army Research Office (ARO) under grant 66270-CS.


\bibliographystyle{IEEEtran}

%

\end{document}